\documentclass[aip,jcp,reprint,amsmath,amssymb]{revtex4-2}
\usepackage{graphicx}
\usepackage[update,prepend]{epstopdf}
\newcommand{\dd}{\mathop{}\!\mathrm{d}}
\newcommand{\vv}{\mathbf}
\newcommand{\vvg}{\boldsymbol}

\usepackage{natmove}

\begin{document}

\title{Effects of linker flexibility on phase behavior and structure of linked colloidal gels}
\author{Michael P. Howard}
\affiliation{McKetta Department of Chemical Engineering, University of Texas at Austin, Austin, Texas 78712, USA}

\author{Zachary M. Sherman}
\affiliation{McKetta Department of Chemical Engineering, University of Texas at Austin, Austin, Texas 78712, USA}

\author{Adithya N Sreenivasan}
\affiliation{McKetta Department of Chemical Engineering, University of Texas at Austin, Austin, Texas 78712, USA}

\author{Stephanie A. Valenzuela}
\affiliation{Department of Chemistry, University of Texas at Austin, Austin, Texas 78712, USA}

\author{Eric V. Anslyn}
\affiliation{Department of Chemistry, University of Texas at Austin, Austin, Texas 78712, USA}

\author{Delia J. Milliron}
\affiliation{McKetta Department of Chemical Engineering, University of Texas at Austin, Austin, Texas 78712, USA}

\author{Thomas M. Truskett}
\email{truskett@che.utexas.edu}
\affiliation{McKetta Department of Chemical Engineering, University of Texas at Austin, Austin, Texas 78712, USA}
\affiliation{Department of Physics, University of Texas at Austin, Austin, Texas 78712, USA}

\begin{abstract}
Colloidal nanocrystal gels can be assembled using a difunctional ``linker'' molecule to mediate bonding between nanocrystals. The conditions for gelation and the structure of the gel are controlled macroscopically by the linker concentration and microscopically by the linker's molecular characteristics. Here, we demonstrate using a toy model for a colloid--linker mixture that linker flexibility plays a key role in determining both phase behavior and structure of the mixture. We fix the linker length and systematically vary its bending stiffness to span the flexible, semiflexible, and rigid regimes. At fixed linker concentration, flexible-linker and rigid-linker mixtures phase separate at low colloid volume fractions in agreement with predictions of first-order thermodynamic perturbation theory, but the semiflexible-linker mixtures do not. We correlate and attribute this qualitatively different behavior to undesirable ``loop'' linking motifs that are predicted to be more prevalent for linkers with end-to-end distances commensurate with the locations of chemical bonding sites on the colloids. Linker flexibility also influences the spacing between linked colloids, suggesting strategies to design gels with desired phase behavior, structure, and by extension, structure-dependent properties.
\end{abstract}
\maketitle

\section{Introduction}
Inorganic nanocrystals (NCs) exhibit distinctive optoelectronic properties such as quantum dot emission, localized surface plasmon resonance (LSPR), and electrocatalytic activity that are controlled by their size, shape, and composition \cite{ElSayed2001,ElSayed2004,Yin2005,Talapin2010,Agrawal2018}.
Properties of individual NCs can be further modified in assembled colloidal structures through coupling to neighboring NCs \cite{Boles2016}. Coupling effects, such as shifts in the LSPR and enhancement of the local electric field, are sensitive to the distance between (and number of) adjacent NCs \cite{Talapin2010,Agrawal2018}, so structure gives another powerful route to tune properties of materials made from NCs. Ordered NC structures like superlattices have long been sought \cite{Collier1998} but only a limited number of (typically close-packed) lattices have been realized \cite{Murray2000}. Gels, i.e., disordered solid-like networks of NCs, are interesting alternative assemblies because their open fractal structures can exhibit varying degrees of NC connectivity giving rise to tunable mechanical or optical properties \cite{Arachchige2007B,Ziegler2017,Rechberger2017}. NC gels have been proposed as photovoltaic materials and electrocatalysts \cite{Nozik2010,Gaponik2011,Cai2018}, and the low NC density in gels also makes them attractive candidates for dynamically reconfigurable materials \cite{Lesnyak2011,Borsley2016,Cabezas2018,Wang2019,Marro2020}.

There are various routes to NC gelation \cite{Zaccarelli2007,Matter2020}, but we have recently demonstrated one controllable, reversible strategy based on dynamic covalent chemistry \cite{Dominguez2020}. Metal oxide NCs were functionalized with ligands bearing an aldehyde group that could react with a complementary hydrazide group on a difunctional ``linker'' molecule. In qualitative agreement with theoretical predictions \cite{Lindquist:2016,Howard:2019}, gelation occurred upon addition of a sufficient amount of linker and could be subsequently reversed upon its dilution. Linker-mediated gelation is experimentally convenient because it is macroscopically controlled by the linker concentration rather than microscopically controlled by the NC functionality. Furthermore, the large synthetic toolbox of organic chemistry offers significant tunability via molecular design of the ligands or linkers. For example, the length (molecular weight) of the ligand--linker complex not only constrains the allowable spacing between two linked NCs but also affects the phase behavior of the NC--linker mixture. We have shown for a model NC--linker system that, at fixed linker-to-NC number ratio, longer linkers compress
the unstable region of the phase diagram (where gels can form by arrested spinodal decomposition upon quenching) to lower NC volume fractions \cite{Howard:2019}.

\begin{figure}
    \centering
    \includegraphics{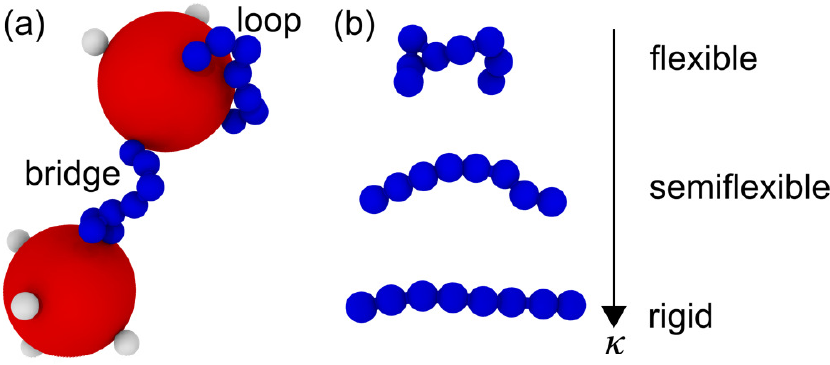}
    \caption{(a) Linkers forming a bridge between two colloids and a loop with the same colloid. (b) Linkers with flexibility spanning the flexible, semiflexible, and rigid regimes (top to bottom: $\beta\kappa = 0$, $7$, and $28$ for the model of Section~\ref{sec:model}). All snapshots were rendered using OVITO 2.9.0 \cite{Stukowski:2010}.}
    \label{fig:schematic}
\end{figure}

Internal molecular degrees of freedom, such as bending stiffness, also play an important role in linker-mediated gelation. We have found that flexible ligands (linkers) can form ``self-links'' or ``loops,'' with difunctional linkers bonding both of their ends to ligands on the same NC (Figure~\ref{fig:schematic}a). When present, such bonding motifs tend to inhibit gelation \cite{Howard:2019,Dominguez2020}. Similar behavior was recently reported in hydrogels made from trivalent DNA nanostars linked by difunctional double-stranded DNA \cite{Stoev:2020}. Double-stranded DNA is a stiff molecule with a relatively large persistence length, but flexibility can be added to a short DNA linker by including single-stranded DNA joints near its ends \cite{Xing:2018}. Double-stranded DNA linkers with flexible joints produced a cluster fluid instead of the expected gel because the linker was able to form a loop with two arms of the same nanostar \cite{Stoev:2020}.

Synthetic ligands and linkers can be designed to have chemical structures that span the flexible, semiflexible, and rigid regimes of bending stiffness (Figure~\ref{fig:schematic}b). Here we highlight a few potential synthetic strategies for realizing such molecules. Flexible molecules, e.g., alkyl chains, are readily made by a variety of techniques; in Ref.~\citenum{Dominguez2020}, we designed and synthesized a flexible ligand for our NCs under amide coupling conditions that are commonly used for preparing natural and unnatural polypeptides \cite{Chen:2013,Wriggers:2005}. At the other extreme, rigid (rodlike) molecules can be realized using aromatic heterocycles with extended delocalization of $\pi$-elections; for example, poly[benzol,2-d:5,4-d'bisoxazole-2,6-diyl)-1,4-phenylene] (PBO) and poly[benzo 1,2-d:4,5-d'bisthiazole-2,6-diyl)-1,4-phenylene] (PBZT) are rigid polymers consisting of conjugated benzobisazole and benzobisthiazole rings, respectively. Rigid polymers can be prepared under polymerization condensation conditions \cite{Wolfe:1981,Hu:2003}, but bonds between aryl rings will also form under copper-, nickel-, or palladium-catalyzed cross-coupling conditions,\cite{Ullmann:1901,Percec:1995,Yun:2020,Nelson:2004} including Kumada \cite{Tamao:1972}, Negishi \cite{Negishi:1977}, Stille \cite{McKean:1987}, and Suzuki \cite{Miyaura:1995} coupling schemes. Semiflexible molecules, with stiffness intermediate between flexible and rigid molecules, can be produced by introducing flexible alkyl chains like polyethylene glycol between rigid, planar aryl moieties under similar synthetic conditions used for rigid molecules \cite{Kikutani:2001,Morozova:2018}. Techniques such as Hiyama-coupling conditions \cite{Hatanaka:1988} can also incorporate methylene groups between aryl motifs \cite{Denmark:2005,McLaughlin:2015,Miller:2014,Zhang:2014,Yoshiaki:2007,Dey:2008}. The length of the flexible alkyl chains or number of methylene groups between aryl motifs can be used to make the molecule more flexible or rigid.

The large space of potential ligand and linker designs with varying molecular weight and flexibility promises significant tunability, but little is known theoretically about how flexibility might affect the phase behavior and structure of linker-mediated NC gel assemblies. To address this gap, here we perform a fundamental study using a toy colloid--linker model (Section~\ref{sec:model}). Building on recent work \cite{Howard:2019}, we systematically introduce bending stiffness to the linkers and show (Section~\ref{sec:results}) that the phase behavior is highly sensitive to the number of loops that form; phase separation (gelation) is suppressed by large loop fractions at fixed colloid and linker concentrations. Linker flexibility also systematically alters the gel structure (and by extension structure-influenced properties), suggesting further avenues for engineering linker-mediated NC gelation (Section~\ref{sec:conclude}).

\section{Model and Simulation Methods}
\label{sec:model}
As in our previous work \cite{Howard:2019}, we studied a colloid--linker mixture having $N_{\rm c}$ colloids and $N_{\rm l}$ linkers in a volume $V$ at temperature $T$. To focus on the effects of flexibility, we considered linkers that were all linear chains having $M = 8$ beads of diameter $d$ and mass $m$. The colloids were modeled as larger beads with diameter $d_{\rm c} = 5\,d$ and mass $m_{\rm c} = 125\,m$. The linker-to-colloid number ratio $\Gamma = N_{\rm l}/N_{\rm c}$ of the mixture was set to $\Gamma = 1.5$, and different values of the colloid volume fraction $\eta_{\rm c} = N_{\rm c} \pi d_{\rm c}^3/(6 V)$ were considered.

All particles were treated as nearly hard spheres interacting through the core-shifted Weeks-Chandler-Andersen potential \cite{Weeks:1971},
\begin{equation}
    \beta u_{\rm r}(r) = \begin{cases}
    4 \left[\left(\dfrac{d}{r-\delta_{ij}} \right)^{12} - \left(\dfrac{d}{r-\delta_{ij}} \right)^6 \right] + 1, & r \le d_{ij}^* \\
    0, & r > d_{ij}^*
    \end{cases},
    \label{eq:wca}
\end{equation}
where $\beta = (k_{\rm B} T)^{-1}$, $k_{\rm B}$ is Boltzmann's constant, $r$ is the distance between the centers of two particles of types $i$ and $j$, and $\delta_{ij} = (d_i + d_j)/2 - d$ shifts the divergence of the potential to account for the diameters, $d_i$ and $d_j$, of each particle type. The cutoff for the potential was chosen as $d_{ij}^* = 2^{1/6} d + \delta_{ij}$ to give a purely repulsive interaction.

Within a linker, pairs of bonded beads were additionally joined by finitely extensible nonlinear elastic springs \cite{Bishop:1979},
\begin{equation}
    u_{\rm s}(r) = \begin{cases}
    -\dfrac{k_{\rm s} r_{\rm s}^2}{2} \ln\left[1 - \left(\dfrac{r}{r_{\rm s}} \right)^2 \right], & r < r_{\rm s} \\
    \infty, & r \ge r_{\rm s}
    \end{cases}.
\end{equation}
Standard parameter choices $k_{\rm s} = 30\,k_{\rm B}T/d^2$ and $r_{\rm s} = 1.5\,d$ were adopted \cite{Grest:1986}, and the average linker bond length was $b \approx 0.97\,d$. The linker flexibility was controlled by including a bending potential between sets of three consecutively bonded beads \cite{Nikoubashman:2016,Nikoubashman:2017},
\begin{equation}
u_\theta(\theta) = \kappa(1+\cos\theta),
\end{equation}
where $\theta$ is the angle between the beads and $\kappa$ sets the bending stiffness. For $\beta\kappa > 2$, the persistence length can be estimated as $\ell \approx \beta \kappa d$ \cite{Nikoubashman:2016}, which should be compared to the average contour length $L = (M-1)b$. The linker is considered \textit{flexible} when $\ell \ll L$, \textit{semiflexible} when $\ell \approx L$, and \textit{rigid} when $\ell \gg L$. We studied bending stiffnesses $0 \le \beta \kappa \le 28$ to span these regimes, giving $\ell/L \lesssim 4$.

Chemical bonding between functional end groups on the linkers with specific bonding sites on the colloids was modeled by fixing six beads (each having nominal mass $m$ and diameter $d$) on the surface of each colloid at radius $d_{\rm cl}^* \approx 3.12\,d$. The end beads of the linkers had a short-ranged attraction $u_{\rm b}$ to the colloid bonding sites,
\begin{equation}
    u_{\rm b}(r) = \begin{cases}
    -\varepsilon \exp\left[-\left(\dfrac{r}{0.2\,d}\right)^2\right], & r \le 0.5\,d \\
    0, & r > 0.5\,d
    \end{cases},
    \label{eq:ub}
\end{equation}
where $\varepsilon$ sets the bond strength. The large $\beta\varepsilon$ limit corresponds to potential experimental realizations of our model using, e.g., dynamic covalent chemistry to form strong but reversible bonds \cite{Rowan2002,Jin2013,Seifert2016,Dominguez2020}. Consistent with this motivation, the colloid bonding sites also repelled each other via Eq.~\eqref{eq:wca} to ensure that only one site could bond with one linker bead at a time. The sites were initially positioned at the vertices of an octahedron, but we also investigated the effects of other arrangements (Section~\ref{sec:patch}).

We performed classical molecular dynamics simulations of our model using \textsc{lammps} (22 Aug 2018) \cite{Plimpton:1995}. The integration timestep was $0.001\,\tau$, where $\tau = \sqrt{\beta m d^2}$ is the unit of time, and constant temperature was maintained using a Langevin thermostat with friction coefficient $0.1\,m/\tau$ applied to each linker bead and colloid. Random initial configurations of $N_{\rm c} = 1000$ colloids and $N_{\rm l} = 1500$ linkers were prepared in cubic simulation boxes with periodic boundary conditions. The colloid volume fraction was varied from $0.01 \le \eta_{\rm c} \le 0.15$ to focus on the dilute conditions typically relevant for preparing low-density nanocrystal gels \cite{Arachchige2007B,Ziegler2017,Rechberger2017,Nozik2010,Gaponik2011,Cai2018}. For a given bending stiffness $\kappa$, the configurations were first equilibrated for $0.5 \times 10^4\,\tau$ without any attraction ($\beta\varepsilon = 0$). Next, the attraction was slowly switched on to $\beta\varepsilon = 10$ over a period of $10^4\,\tau$ using a linear ramp, followed by a $10^4\,\tau$ equilibration period. We then linearly ramped the attraction to $\beta\varepsilon = 11$ over a period of $0.5 \times 10^4\,\tau$ and held it constant for $1.5 \times 10^4\,\tau$. We saved the configuration from the end of this period and repeated the incrementing process until $\beta\varepsilon = 20$. Finally, production simulations of $10^4\,\tau$ were performed at each $\eta_{\rm c}$ and $\varepsilon$ with configurations sampled every $10\,\tau$ for analysis.

In order to determine approximate phase boundaries for the colloid--linker mixture, which give guidance about conditions that may lead to gelation, we computed the partial static structure factor of the colloids \cite{Hansen2013},
\begin{equation}
    S_{\rm cc}(\vv{q}) = \frac{1}{N_{\rm c}} \left\langle \sum_{j,k}^{N_{\rm c}} e^{-i \vv{q}\cdot(\vv{r}_j-\vv{r}_k)} \right\rangle.
\end{equation}
Here $\vv{q} = (2\pi/V^{1/3})\vv{n}$ is a wavevector in the cubic box of volume $V$, $\vv{n}$ is a vector of integers, and $\vv{r}_j$ is the position of the $j$th colloid. The angle brackets indicate an ensemble average, which was computed as a time average.  We determined the average $S_{\rm cc}(q)$ for the 22 smallest nonzero wavevector magnitudes $q=|\vv{q}|$ and extrapolated to $q=0$ by fitting a Lorentzian form \cite{Jadrich:2015},
\begin{equation}
    S_{\rm cc}(q) \approx \frac{S_{\rm cc}(0)}{1-(q\xi_{\rm cc})^2},
\end{equation}
where $\xi_{\rm cc}$ is a correlation length. We operationally defined structures having $S_{\rm cc}(0) > 10$ to be phase separated \cite{Zaccarelli:2005,Lindquist:2016,Howard:2019}.

\section{Results and Discussion}
\label{sec:results}

\subsection{Phase behavior}
The phase behavior of a dilute mixture of colloids and semiflexible or rigid linkers is expected to be qualitatively similar to that of mixtures of patchy colloids \cite{Lindquist:2016} or colloids and flexible linkers \cite{Howard:2019}. The homogeneous colloid--linker mixture is typically thermodynamically stable when the attraction between the colloids and linkers is weak (small $\beta\varepsilon$) but can become unstable to composition fluctuations when the attraction is strong (large $\beta\varepsilon$) and enough linker is added (increasing $\Gamma$), leading to separation into colloid-rich and colloid-lean phases. We note that the stable fluid phase can be reentrant with respect to $\Gamma$ \cite{Lindquist:2016,Howard:2019}, but our simulations are performed in the linker-limited regime where increasing $\Gamma$ destabilizes the fluid.

The thermodynamically unstable regions of the phase diagram are demarcated by the spinodal boundary \cite{Hansen2013,Chimowitz:2005}. Mixtures prepared inside this region will spontaneously phase separate by spinodal decomposition, and this process typically produces kinetically arrested gels \cite{Zaccarelli2007}. Mixtures prepared outside both the spinodal region and the binodal region (demarcating true phase coexistence) remain single phase but can form gels if the lifetime of the linker-mediated connections between colloids becomes long compared to the observation time \cite{Bianchi:2006,Zaccarelli2007}. Because of their inherent thermodynamic stability, these ``equilibrium gels'' are more resistant to aging than gels formed by arrested spinodal decomposition. Theoretical knowledge of the phase diagram can guide design of materials that controllably gel by one these routes.

We previously applied Wertheim's first-order thermodynamic perturbation theory (TPT) \cite{Wertheim:1984a,Wertheim:1984b,Wertheim:1986a,Wertheim:1986b,Chapman:1988} to predict the conditions for phase separation in a mixture of colloids and fully flexible ($\beta\kappa = 0$) linkers \cite{Howard:2019}. Within TPT, the Helmholtz free-energy density $a$ is decomposed into two contributions, $a = a_0 + a_{\rm b}$. $a_0$ is the free-energy density of a reference system without bonding between components, and $a_{\rm b}$ is the free-energy density due to bonding. For fully flexible linkers, the hard-chain fluid \cite{Chapman:1988} is a useful reference system and has
\begin{equation}
    \beta a_0 = \beta a_{\rm id} + \beta a_{\rm hs}^{\rm ex} - \sum_i \rho_i (M_i-1) \ln g_{\rm hs}^{(ii)}(d_i^+).
    \label{eq:ahc}
\end{equation}
The first term is the free-energy density of an ideal gas of chain molecules,
\begin{equation}
    \beta a_{\rm id} = \sum_i \rho_i [\ln (\Lambda_i^3 \rho_i)-1],
\end{equation}
where the sum is taken over the different components $i$, each of which consists of chains of $M_i$ beads of equal diameter $d_i$. Here, $\rho_i$ is the number density, and $\Lambda_i$ is the thermal wavelength of component $i$. For our model, there are two components: colloids each having $M_{\rm c} = 1$ bead of diameter $d_{\rm c}$ and with $\rho_{\rm c} = 6 \eta_{\rm c}/(\pi d_{\rm c}^3)$, and linkers having $M_{\rm l} = M$ beads of diameter $d$ and with $\rho_{\rm l} = \Gamma\rho_{\rm c}$. The second term in Eq.~\eqref{eq:ahc}, $a_{\rm hs}^{\rm ex}$, is the excess free-energy density of the hard-sphere fluid obtained by dissociating all bonds from the chains \cite{Boublik:1970,Howard:2017},
\begin{equation}
    \beta a_{\rm hs}^{\rm ex} = \frac{6}{\pi}\left[\left(\frac{\xi_2^3}{\xi_3^2}-\xi_0\right)\ln(1-\xi_3) + \frac{3\xi_1\xi_2}{1-\xi_3} + \frac{\xi_2^3}{\xi_3(1-\xi_3)^2}\right],
\end{equation}
where $\xi_j = \sum_i \rho_i M_i \pi d_i^j/6$. The third term is the excess free-energy density of chain formation \cite{Chapman:1988}, where $g_{\rm hs}^{(ii)}(d_i^+)$ is the contact value of the radial distribution function for two beads of component $i$ in the hard-sphere fluid \cite{Boublik:1970}. Note that Eq.~\eqref{eq:ahc} corrects a typographical error in Eq.~(5) of Ref.~\citenum{Howard:2019}, where the ideal gas and excess hard-sphere free energies were incorrectly described; Ref.~\citenum{Howard:2019} in fact used Eq.~\eqref{eq:ahc} to perform calculations.

To compute the bonding free-energy density $a_{\rm b}$, it is usually assumed that to first order each site can bond with at most one other site, that bonding is uncorrelated between sites, and that only treelike bonded networks form \cite{Jackson:1988}. For a mixture of components each having $n_i$ identical bonding sites, $a_{\rm b}$ can be expressed as \cite{Chapman:1988}
\begin{equation}
    \beta a_{\rm b} = \sum_i \rho_i n_i \left[ \ln X_i + \frac{1}{2}(1-X_i) \right],
    \label{eq:ab}
\end{equation}
where $X_i$ is the fraction of sites not bonded on component $i$. In our model, the colloids have $n_{\rm c} = 6$ sites and the linkers have $n_{\rm l} = 2$ sites. The values of $X_i$ are determined by the chemical equilibrium equations,
\begin{equation}
    X_i = \bigg(1 + \sum_j \rho_j n_j X_j \Delta_{ij} \bigg)^{-1},
\end{equation}
where only $\Delta_{\rm cl}$ is nonzero for our model and is given by \cite{Howard:2020}
\begin{equation}
    \Delta_{\rm cl} = \frac{1}{8\pi^2} \int \dd\vv{r} \dd\vvg{\Omega} \dd\vv{R}~p_0(\vv{R}) g_0^{({\rm cl})}(\vv{r},\vv{R}) f^{({\rm cl})}(\vv{r},\vvg{\Omega},\vv{R}).
    \label{eq:dcp}
\end{equation}
Here, $\vv{r}$ is the vector from the center of the colloid to the end of the linker participating in the bond, $\vvg{\Omega}$ is the vector of Euler angles defining the orientation of the colloid, and $\vv{R}$ is the end-to-end vector of the linker. The integrand is the product of the distribution of end-to-end vectors $p_0(\vv{R})$ in the hard-chain reference fluid, the colloid--linker pair correlation function $g_0^{({\rm cl})}$ in the hard-chain reference fluid, and the Mayer $f$-function $f^{({\rm cl})}$ for a designated bonding site on the colloid and the end of the linker at $\vv{r}$. The $f$-function, and hence the integrand, is nonzero only when the two sites interact, which occurs over a short range given the form of Eq.~\eqref{eq:ub}. $\Delta_{\rm cl}$ can be regarded as a ``bond volume'' averaged over orientations of the colloid and conformations of the linker. We evaluated $\Delta_{\rm cl}$ using the same methodology as in Ref.~\citenum{Howard:2020} \cite{numpy,scipy,numba}, where $p_0$ and $g_0^{({\rm cl})}$ were obtained from simulations of the hard-chain reference fluid; complete details are given in the Supporting Information.

\begin{figure*}
    \centering
    \includegraphics{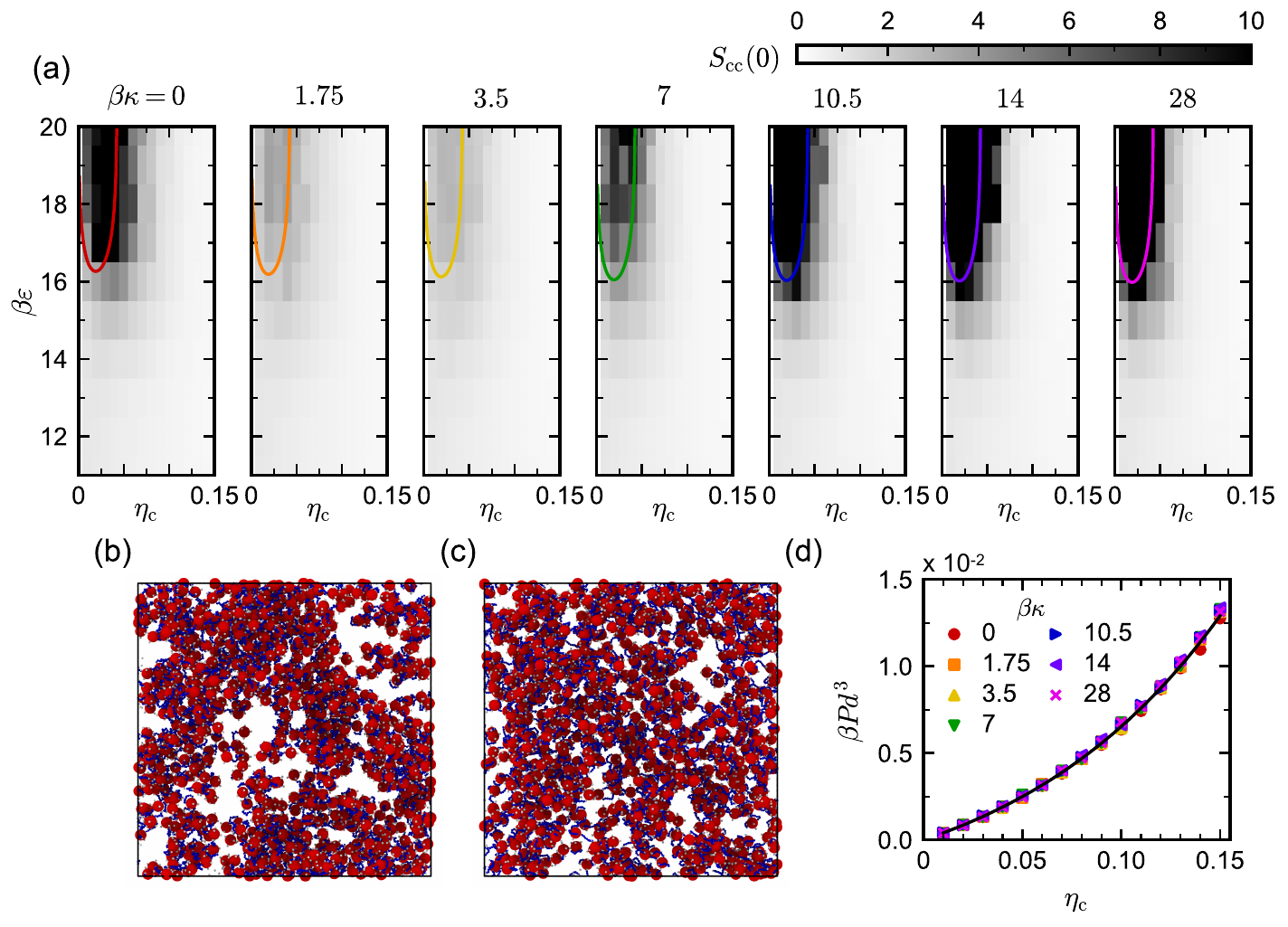}
    \caption{(a) Phase diagrams for colloids mixed with linkers of varying bending stiffness $\kappa$ in the $(\eta_{\rm c},\varepsilon)$ plane at $\Gamma = 1.5$. The lines show the spinodal boundary computed from first-order TPT, and the grayscale map gives the colloid partial structure factor extrapolated to zero-wavevector, $S_{\rm cc}(0)$, from the simulations. Large values of $S_{\rm cc}(0)$ indicate phase separation in the simulations; usually, $S_{\rm cc}(0) > 10$ is considered separated. Snapshots are shown at $\eta_{\rm c} = 0.03$ and $\beta\varepsilon = 20$ for (b) a nonuniform (phase-separated) morphology when $\beta\kappa = 0$ and (c) a more uniform morphology when $\beta\kappa = 3.5$. (d) The pressure $P$ of the reference fluid without linking was similar for all $\kappa$ and in good agreement with the theoretical pressure of the hard-chain fluid (black line).}
    \label{fig:phase}
\end{figure*}

Given the Helmholtz free energy $a$, the spinodal boundary can be computed using the matrix $\vv{H}$ of second derivatives with respect to density \cite{Chimowitz:2005}, where $H_{ij} = \partial^2 a/\partial \rho_i \partial \rho_j$. The limit of stability occurs when the determinant of $\vv{H}$ is zero. For the fully flexible linkers ($\beta\kappa = 0$), the TPT predictions are in good agreement with the simulations (Figure~\ref{fig:phase}a). (The TPT predictions are also improved compared to Ref.~\citenum{Howard:2019} because Eq.~\eqref{eq:dcp} uses a more accurate approximation of the pair correlation function.) Phase separation was mainly observed in the simulations inside the TPT spinodal; Figure~\ref{fig:phase}b shows a representative phase-separated morphology at $\eta_{\rm c} = 0.03$ and $\beta\varepsilon = 20$. We previously investigated how the phase behavior of mixtures of colloids and fully flexible linkers depended on the linker length \cite{Howard:2019}, so here we focus on the role of flexibility.

\begin{figure}
    \centering
    \includegraphics{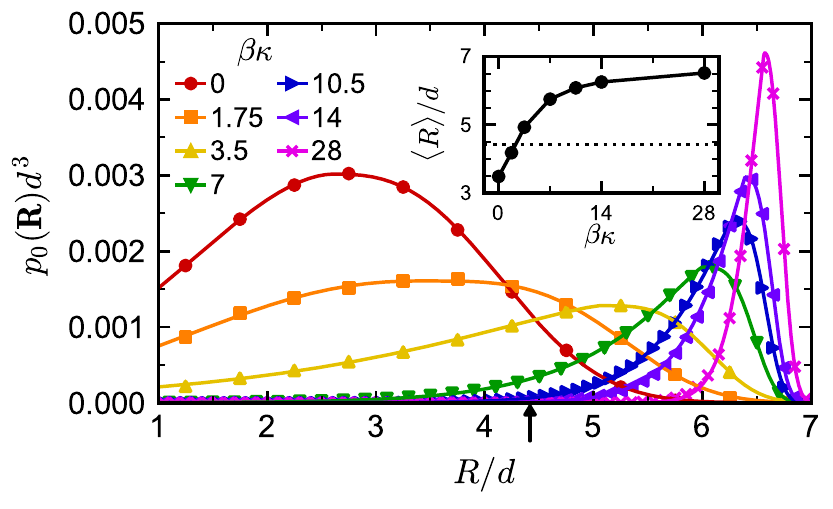}
    \caption{Distribution of the linker end-to-end vector $\vv{R}$ in the reference fluid as a function of the end-to-end distance $R = |\vv{R}|$ at $\eta_{\rm c} = 0.01$, $\Gamma = 1.5$, and different bending stiffnesses $\kappa$. Curves are empirical fits of the distributions used to evaluate Eq.~\eqref{eq:dcp} (Eq.~(S12)). The inset shows the average end-to-end distance $\langle R \rangle$ as a function of $\kappa$. The arrow and dotted line indicate the distance between neighboring vertices of the octahedron, $\sqrt{2} d_{\rm cl}^*$. The normalization of $p_0(\vv{R})$ is such that $\int \dd\vv{R} p_0(\vv{R}) = 1$.}
    \label{fig:dists}
\end{figure}

As a perturbation theory, the TPT free energy is sensitive to the reference fluid model, i.e., without linking. In particular, the free-energy density $a$ depends on both the reference fluid's equation of state, which determines $a_0$, and its structure, which enters $a_{\rm b}$ through $\Delta_{\rm cl}$. We hence expect similar phase behavior for different bending stiffnesses $\kappa$ if the equation of state and relevant structures of the reference fluid do not depend strongly on $\kappa$. We performed simulations of mixtures having $\beta \varepsilon = 0$, $\Gamma = 1.5$, and varied $\eta_{\rm c}$ and $\kappa$, and we measured the pressure $P$ (Figure~\ref{fig:phase}d), the distribution of linker end-to-end vectors $p_0(\vv{R})$ (Figure~\ref{fig:dists}), and the pair distribution function $g_0^{({\rm cl})}$ (Figures S1 and S2). The pressure changed negligibly with $\kappa$ and was fully consistent with the pressure derived from Eq.~\eqref{eq:ahc} \cite{Chapman:1988}. The end-to-end vector distribution $p_0(\vv{R})$ changed significantly with $\kappa$, shifting its peak to larger $R = |\vv{R}|$ for larger $\kappa$ as expected. The pair distribution function $g_0^{({\rm cl})}$ was again far less sensitive to $\kappa$. It tended to be less than 1 for conformations relevant to evaluating Eq.~\eqref{eq:dcp} due to depletion of the polymer-like chains near the surface of the colloid \cite{Fuchs:2002}, with slightly larger values for the stiffer linkers. Overall, though, the change in $\Delta_{\rm cl}$ at $\beta\varepsilon = 20$ was only 30\% between $\beta\kappa = 0$ and $\beta\kappa = 28$.

Given this analysis of the reference fluid, we expected similar phase behavior at all values of $\kappa$. Indeed, for $\beta\kappa \gtrsim 10.5$ (the rigid linkers), the simulated phase diagrams were essentially the same, closely resembling that at $\beta\kappa = 0$ and the TPT predictions. Surprisingly, though, the semiflexible linkers ($1.75 \le \beta\kappa \lesssim 7$) exhibited very different behavior than predicted by first-order TPT. For $\beta\kappa = 1.75$ and $3.5$, only minor structuring was detected at all $\eta_{\rm c}$ and $\varepsilon$, with $S_{\rm cc}(0)$ well below the threshold we considered phase separated. This difference was visually apparent in simulation snapshots taken at conditions inside the TPT spinodal. Unlike when $\beta\kappa = 0$ (Figure~\ref{fig:phase}b), significant colloid density variations were not visible at $\eta_{\rm c} = 0.03$ and $\beta\varepsilon = 20$ when $\beta\kappa = 3.5$ (Figure~\ref{fig:phase}c). The lack of phase separation for only the semiflexible linkers could not be explained using first-order TPT.

\subsection{Linker loops}
We suspected that the striking discrepancy between the simulations and TPT for the semiflexible linkers was caused by certain linking motifs that are prevalent in the simulations at these flexibilities but are not accounted for in first-order TPT. In previous work, we have speculated that the presence of cycles within the graph representing the bonded network of colloids and linkers might suppress phase separation \cite{Howard:2019,Dominguez2020}. In particular, we hypothesized that for a fixed total amount of linker $\Gamma$ the fluid phase might be stabilized if a substantial number of linkers formed ``loops'' by attaching both ends to the same colloid, rather than forming ``bridges'' between two different colloids; a similar explanation was recently proposed for the lack of gelation in DNA hydrogels linked with flexible joints \cite{Stoev:2020}. This definition of a loop, which can also be interpreted as a ``double bond'' between a linker and a colloid, is the smallest cycle that can form, and other cycles like double links between two colloids are included with the bridges here.

\begin{figure}
    \centering
    \includegraphics{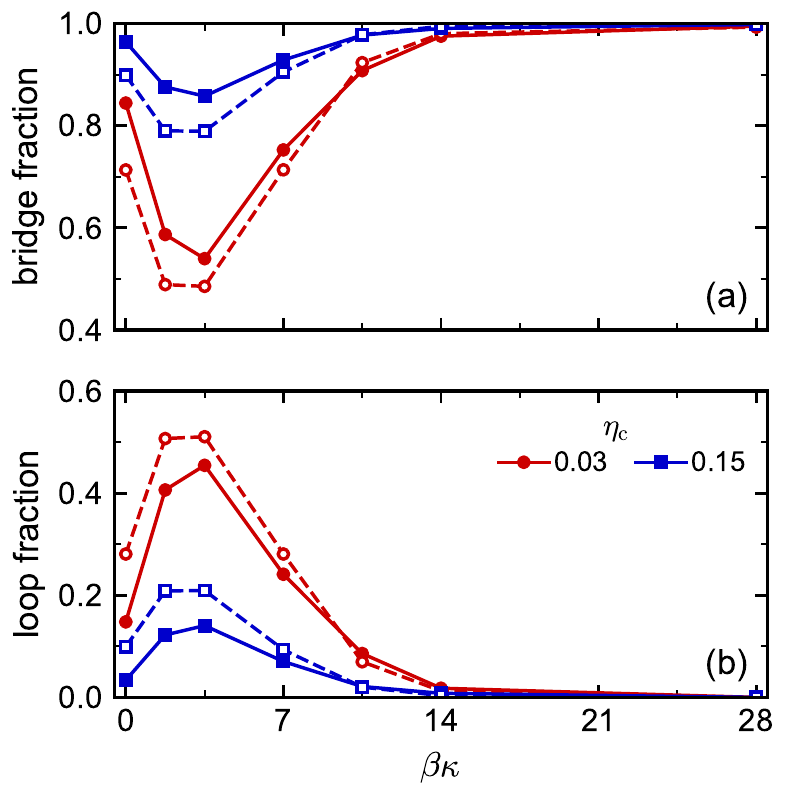}
    \caption{Fraction of linkers forming (a) bridges between colloids and (b) loops to the same colloid in the simulations (filled symbols) and TPT incorporating loops (open symbols) as a function of bending stiffness $\kappa$ at $\beta\varepsilon = 20$ and $\Gamma = 1.5$. Results are shown at two values of $\eta_{\rm c}$, one of which phase separated in some of the simulations ($\eta_{\rm c} = 0.03$) and one of which did not ($\eta_{\rm c} = 0.15$).}
    \label{fig:loops}
\end{figure}

We counted the number of linkers forming either bridges or loops in our simulations (Figure~\ref{fig:loops}), focusing on the strong attraction regime ($\beta\varepsilon = 20$) where essentially all linkers had both ends bonded in one of these motifs. For the fully flexible linkers, a small fraction of linkers formed loops (Figure~\ref{fig:loops}b) rather than bridges (Figure~\ref{fig:loops}a) at all $\eta_{\rm c}$, with the loop fraction tending to be smaller at larger $\eta_{\rm c}$. An overwhelming majority of rigid linkers formed bridges at all $\eta_{\rm c}$, constituting $\ge 90\%$ of the links for $\beta\kappa \ge 14$. Interestingly, and consistent with our hypothesis, there was a pronounced increase in the loop fraction (Figure~\ref{fig:loops}b) for the semiflexible linkers, which had a maximum at roughly $\beta\kappa = 3.5$. At $\eta_{\rm c} = 0.03$, nearly 50\% of the linkers formed loops when $\beta\kappa = 3.5$ and the mixture remained essentially single phase, despite phase separating for both the fully flexible and rigid linkers (Figure~\ref{fig:phase}a).

We recently extended first-order TPT to include linker loops (double bonds) \cite{Howard:2020}, and this theory readily predicts the fraction of linkers forming bridges or loops in a spatially homogeneous fluid at equilibrium. We computed these fractions at $\eta_{\rm c} = 0.03$ and $\eta_{\rm c} = 0.15$ for all chain flexibilities using a similar methodology as in Ref.~\citenum{Howard:2020}; numerical details are given in the Supplementary Material. The TPT predictions for the fraction of linkers forming loops are in excellent agreement with the simulations (Figure~\ref{fig:loops}), supporting the notions that (1) loops are equilibrium bonding motifs and (2) for the studied colloid bonding site geometry and linker length, loops are more prevalent at intermediate flexibilities. We note that the TPT neglects any phase separation in the simulations at $\eta_{\rm c} = 0.03$; this does not seem to qualitatively affect the result.

The maximum in loop fraction with respect to $\kappa$ can be understood using the distribution of linker end-to-end vectors (Figure~\ref{fig:dists}). In order to form a loop, both ends of the linker must attach to sites on the same colloid, so loops become more prevalent when it is more probable for the linker to have an end-to-end distance commensurate with the colloid site--site distance. We consider only loops between patches at neighboring vertices of the octahedron, which are separated by distance  $\sqrt{2} d_{\rm cl}^*$ (arrow in Figure~\ref{fig:dists}), because loops between vertices on opposite hemispheres of the colloid are forbidden by the linker contour length. As $\kappa$ increases, this end-to-end distance initially becomes more probable but then becomes significantly less likely once the linkers become rigid. The average end-to-end distance $\langle R \rangle$ (inset of Figure~\ref{fig:dists}) is also comparable to the site--site distance (dotted line) in the semiflexible regime where the most loops are observed. Qualitatively, the flexible linkers pay an entropic penalty to stretch beyond their preferred size and form a loop, while the rigid linkers pay a bending-energy penalty to do the same; in contrast, the semiflexible linkers are naturally compatible with the length scale required to form loops in our model. These effects of flexibility measured by $p_0$ manifest in the TPT as an increase in the double-bond volume \cite{Howard:2020}, which is nearly three times larger when $\beta\kappa = 3.5$ than when $\beta\kappa = 0$.

\subsection{Colloid bonding site geometry}
\label{sec:patch}
\begin{figure}
    \centering
    \includegraphics{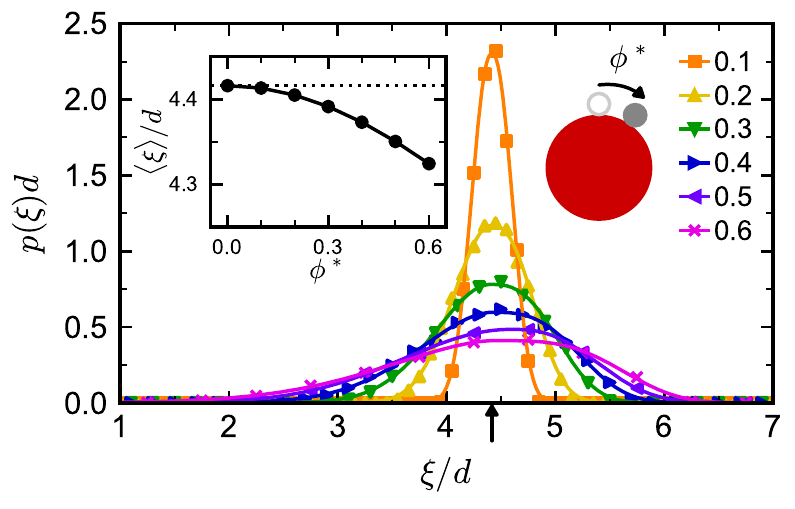}
    \caption{Distribution of the distance $\xi$ between pairs of colloid bonding sites that were initially nearest neighbors. The sites were perturbed by a displacement drawn uniformly using spherical coordinates with a maximum polar rangle $\phi^*$. The inset shows the average of $\xi$ as a function of $\phi^*$. The normalization of $p(\xi)$ is such that $\int \dd\xi p(\xi) = 1$.}
    \label{fig:patchdists}
\end{figure}

An immediate consequence of this interpretation is that the loop fraction should also depend on the distribution of distances $\xi$ between bonding sites on the colloids. A distribution $p(\xi)$ that overlaps significantly with $p_0(\vv{R})$ should favor loop formation, while an incompatible $p(\xi)$ should produce fewer loops. The impact of the colloid bonding site geometry on loop formation is not only of theoretical interest but also a practical consideration for experiments, where $p(\xi)$ might be determined by the details of the surface functionalization and polydispersity in surface characteristics is inherent to synthesis. To test the sensitivity of properties to the bonding site geometry, we randomly displaced the colloid bonding sites from their initial positions at the vertices of an octahedron. As illustrated in Figure~\ref{fig:patchdists}, each site was pertubed by a uniformly random amount on the surface of the sphere with radius $d_{\rm cl}^*$ up to a maximum polar angle $\phi^*$. Adjusting $\phi^*$ varies $p(\xi)$ from a perfectly monodisperse distribution at $\phi^* = 0$ to a broad distribution at $\phi^* = 0.6$ (Figure~\ref{fig:patchdists}), which is close to the maximum polar angle that still guarantees two sites will not overlap. (We include in $p(\xi)$ only the distances between sites that were initially nearest neighbors on the octahedron, as these are the ones most likely to form loops even after being displaced.) Generating bonding site polydispersity in this way ensures that the average site--site distance $\langle \xi \rangle$ remains nearly constant (decreasing only 2\% over the entire $\phi^*$ range) and allows us to isolate the effects of site polydispersity from effects of the average site--site distance.

\begin{figure}
    \centering
    \includegraphics{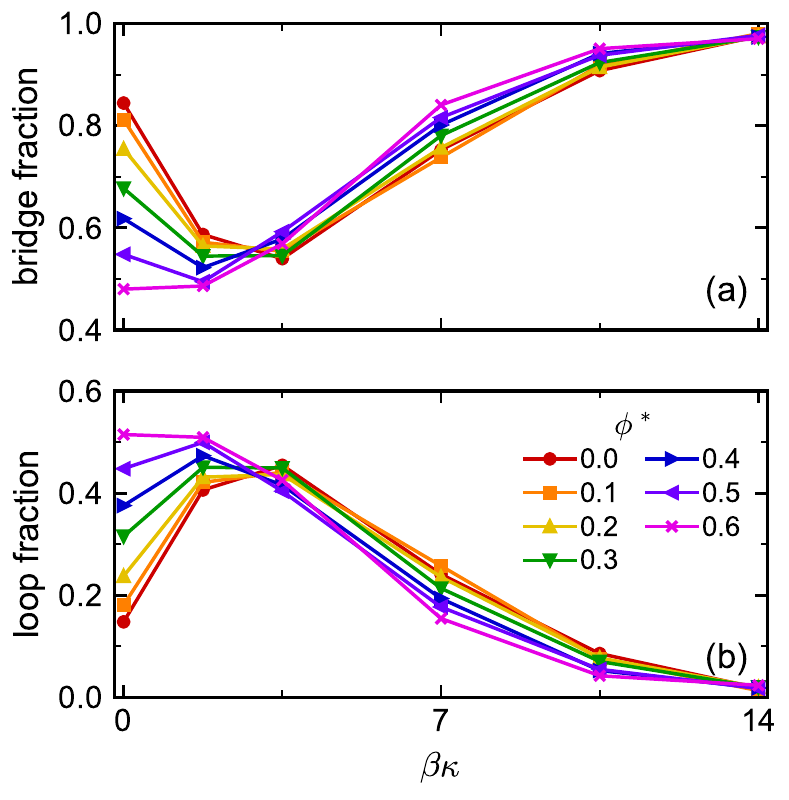}
    \caption{Fraction of linkers forming (a) bridges between colloids and (b) loops to the same colloid in the simulations as a function of bending stiffness $\kappa$ with varied maximum bonding-site displacement $\phi^*$ at $\beta\varepsilon = 20$, $\eta_{\rm c} = 0.03$, and $\Gamma = 1.5$.}
    \label{fig:loopsrandom}
\end{figure}

We simulated the colloid--linker mixture with polydisperse bonding sites at $\eta_{\rm c} = 0.03$, which is inside the first-order TPT spinodal and led to phase separation for the flexible and rigid linkers. We will again focus on the loop fraction and phase behavior in the strong attraction limit $\beta\varepsilon = 20$. For $\phi^* \le 0.2$, the bridge and loop fractions were nearly unchanged from the monodisperse ($\phi^* = 0$) case, with the loop fraction having a maximum for the semiflexible linkers (Figure~\ref{fig:loopsrandom}). Further increase in $\phi^*$ significantly increased the loop fraction for the flexible linkers, but the rigid linkers were less affected. It is sensible that increasing $\phi^*$ had a more dramatic effect on the flexible linkers than the rigid linkers because $p_0(\vv{R})$ for the flexible linkers is broad (Figure~\ref{fig:dists}); perturbing the colloid bonding sites brings some sites closer together (Figure~\ref{fig:patchdists}) where there are probable flexible-linker conformations that previously did not produce loops. For the rigid linkers, increasing $\phi^*$ did little to increase the overlap between the two distributions, and the loop fraction correspondingly changed little with $\phi^*$.

\begin{figure}
    \centering
    \includegraphics{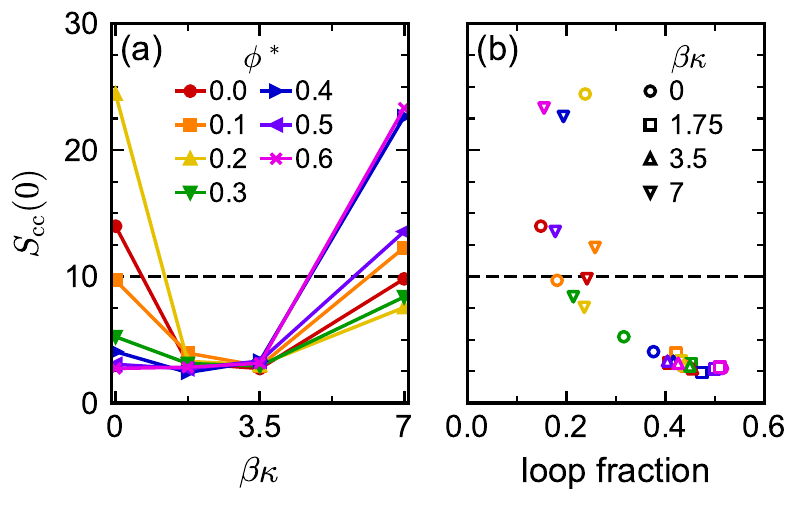}
    \caption{(a) Colloid partial structure factor $S_{\rm cc}$ extrapolated to zero-wavevector as a function of bending stiffness $\kappa$ for varied maximum bonding-site displacement $\phi^*$ at the same state point as in Figure~\ref{fig:loopsrandom}. The dashed line indicates $S_{\rm cc}(0) = 10$, above which we consider the dispersion to be phase separated, and only results for $\beta\kappa \le 7$ are shown because $S_{\rm cc}(0) > 30$ for all $\phi^*$. (b) $S_{\rm cc}(0)$ as a function of loop fraction (Figure~\ref{fig:loopsrandom}b); colors correspond to different values of $\phi^*$ in (a), and symbols designate different $\kappa$.}
    \label{fig:s0kappa}
\end{figure}

We also measured the colloid partial structure factor $S_{\rm cc}(0)$ (Figure~\ref{fig:s0kappa}a) to interrogate how the thermodynamic stability of the fluid phase depended on $\phi^*$. We show only the results for $\beta\kappa \le 7$ because all rigid-linker mixtures phase separated regardless of $\phi^*$. Based on the measured loop fractions, we expected and observed similar phase behavior when $\phi^* \le 0.2$ as for the monodisperse ($\phi^* = 0$) bonding site geometry, with $S_{\rm cc}(0)$ consistent with a lack of phase separation in the semiflexible-linker mixtures. For larger $\phi^*$, phase separation was also suppressed for the flexible linkers, which had a significant increase in the loop fraction. Indeed, we found a strong negative correlation between $S_{\rm cc}(0)$ and the loop fraction (Figure~\ref{fig:s0kappa}b); $S_{\rm cc}(0)$ indicated phase separation when less than 30\% of linkers formed loops. This strongly suggests that controlling the loop fraction, e.g., by designing the linker length or flexibility, is a key component of engineering the phase behavior of the colloid--linker mixture and the conditions that may lead to gelation.

\subsection{Structure}
\label{sec:struct}
So far, we have focused our discussion on controlling phase behavior, but in our previous work with flexible linkers, we showed that adjusting the linker length also gave control over the microstructure of the assembled colloids \cite{Howard:2019}. Longer linkers having larger $\langle R \rangle$ tended to increase the spacing between colloids, which may have important consequences for the optical properties of gels containing plasmonic nanocrystals \cite{Talapin2010,Agrawal2018}. Linker flexibility gives another potential handle to control the colloid microstructure because semiflexible or rigid linkers have larger $\langle R \rangle$ than flexible linkers of equal contour length (Figure~\ref{fig:dists}).

\begin{figure}
    \centering
    \includegraphics{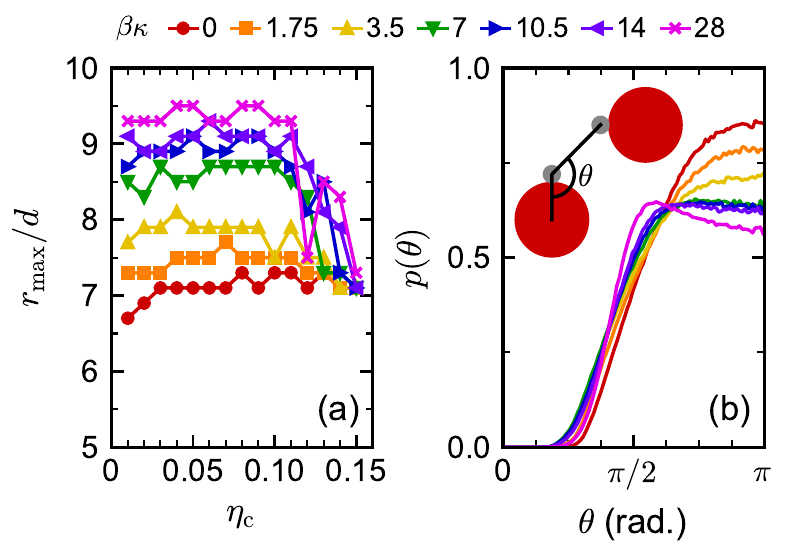}
    \caption{(a) Radial distance $r_{\rm max}$ of maximum value of $g^{({\rm cc})}$ as a function of $\eta_{\rm c}$ for different bending stiffnesses $\kappa$ at $\beta\varepsilon = 20$ and $\Gamma = 1.5$. (b) Distribution of linker bond angles $\theta$ for different bending stiffnesses $\kappa$ at $\beta\varepsilon = 20$, $\eta_{\rm c} = 0.15$, and $\Gamma = 1.5$. The normalization of $p(\theta)$ is such that $\int_0^\pi {\rm d}\theta \sin \theta p(\theta) = 1$.
    }
    \label{fig:rdfpeak}
\end{figure}

We measured the colloid--colloid pair correlation function $g^{({\rm cc})}$ (Figure S3) for various compositions and flexibilities in the strong attraction limit ($\beta\varepsilon = 20$). We focused on the position $r_{\rm max}$ of the maximum in $g^{({\rm cc})}$ (Figure~\ref{fig:rdfpeak}a), which identifies the typical distance to a colloid's nearest neighbor. At small $\eta_{\rm c}$, $r_{\rm max}$ shifted to larger distances with increasing $\kappa$, which is consistent with the corresponding increase in $\langle R \rangle$. Interestingly, this shift did not depend on whether the mixture phase separated, and $r_{\rm max}$ was nearly constant for a given $\kappa$ when $\eta_{\rm c} \lesssim 0.10$. At larger $\eta_{\rm c}$, $r_{\rm max}$ decreased with increasing $\eta_c$ and ultimately reached nearly the same value for all $\kappa$ at $\eta_c = 0.15$. The flexible linker seems to be an exception, having nearly constant $r_{\rm max}$ over the entire range of $\eta_c$ we simulated.

Due to many-body effects in the linked-colloid assemblies, the measured $r_{\rm max}$ cannot be simply explained using only the average end-to-end size of the linker $\langle R \rangle$. Assuming a colinear arrangement of two colloids and a linker, we might expect $r_{\rm max} \approx 2 d_{\rm cl}^* + \langle R \rangle$, which is $9.7\,d$ and $12.7\,d$ for $\beta \kappa = 0$ and $28$, respectively; both are considerably larger than the measured $r_{\rm max}$. To obtain a smaller $r_{\rm max}$, the linkers must either compress or connect two colloids at an angle with each other. A significant decrease in the end-to-end distance of the rigid linkers would incur a large bending penalty and is unlikely, so we accordingly measured the bond angle $\theta$ between pairs of linked colloids (Figure~\ref{fig:rdfpeak}b). We defined $\theta$ using the center and bonded patch of the first colloid and the bonded patch on the second colloid. The distribution of bond angles showed predominantly colinear arrangements ($\theta = \pi$) for the flexible linkers, but as $\kappa$ increased, the bond angles were forced toward more right-angle ($\theta = \pi/2$) arrangements. This change in $\theta$ allows the colloids to pack closely without compressing the linker and seems to supress formation of a single dominant length scale in the assemblies, producing a more uniform distribution of pair distances (Figure S3).

\section{Conclusions}
\label{sec:conclude}
We have shown that flexibility can play an important role in determining the phase behavior and structure of linked colloidal assemblies. In our simulations of a toy model for a colloid--linker mixture, phase separation, which can lead to gelation, was suppressed when the linker's internal degrees of freedom allowed it to readily attach both of its ends to the same colloid in a ``loop.'' This finding is fully consistent with recent experiments with trivalent DNA nanostars \cite{Stoev:2020}, despite the significantly higher number of potential bonding sites on the colloids we studied. The linker loop fraction can be predicted theoretically using an extension of thermodynamic perturbation theory \cite{Howard:2020}, and more efficient linking may then be obtained by designing linker molecules with size, flexibility, or functionality that disfavor or disallow loop motifs. Linker flexibility also gives a handle for controlling the microstructure of the assembled colloids, particularly at small $\eta_{\rm c}$; however, changes in the size of the linker with flexibility do not straightforwardly propagate to the typical colloid--colloid separation, especially at larger $\eta_{\rm c}$. Since many nanoparticle gels are prepared at relatively low volume fractions, flexibility may still be a viable strategy for controlling both gel microstructure and phase behavior. Assembly schemes that use mixtures of linkers having different lengths as well as different flexibilities are especially intriguing, as they may give independent tunability of phase behavior and microstructure \cite{Howard:2019}.

\section*{Supplementary Material}
See supplementary material for details of the TPT calculations including loops, simulated reference fluid data, and colloid--colloid pair correlation functions in linked assemblies.

\begin{acknowledgements}
This research was primarily supported by the National Science Foundation through the Center for Dynamics and Control of Materials: an NSF MRSEC under Cooperative Agreement No.~DMR-1720595, with additional support from an Arnold O. Beckman Postdoctoral Fellowship (ZMS), a National Science Foundation Graduate Research Fellowship under Grant No.~DGE-1610403 (SAV), and the Welch Foundation (Grant Nos.~F-1696 and F-1848). We acknowledge the Texas Advanced Computing Center (TACC) at The University of Texas at Austin for providing HPC resources.
\end{acknowledgements}

\section*{Data Availability}
The data that support the findings of this study are available from the authors upon reasonable request.

\section*{References}
\bibliography{ms}

\end{document}


\title{Supplementary material for ``Effects of linker flexibility on phase behavior and structure of linked colloidal gels''}
\author{Michael P. Howard}
\affiliation{McKetta Department of Chemical Engineering, University of Texas at Austin, Austin, Texas 78712, USA}

\author{Zachary M. Sherman}
\affiliation{McKetta Department of Chemical Engineering, University of Texas at Austin, Austin, Texas 78712, USA}

\author{Adithya N Sreenivasan}
\affiliation{McKetta Department of Chemical Engineering, University of Texas at Austin, Austin, Texas 78712, USA}

\author{Stephanie A. Valenzuela}
\affiliation{Department of Chemistry, University of Texas at Austin, Austin, Texas 78712, USA}

\author{Eric V. Anslyn}
\affiliation{Department of Chemistry, University of Texas at Austin, Austin, Texas 78712, USA}

\author{Delia J. Milliron}
\affiliation{McKetta Department of Chemical Engineering, University of Texas at Austin, Austin, Texas 78712, USA}

\author{Thomas M. Truskett}
\email{truskett@che.utexas.edu}
\affiliation{McKetta Department of Chemical Engineering, University of Texas at Austin, Austin, Texas 78712, USA}
\affiliation{Department of Physics, University of Texas at Austin, Austin, Texas 78712, USA}

\maketitle

We recently derived a TPT incorporating loop (one linker with both ends connected to one colloid) and ring (two linkers connecting two colloids) bonding motifs \cite{Howard:2020}. We included these motifs in a graphical expansion of the free energy and showed the theory accurately predicted the fractions of flexible linkers forming loops and rings in simulations. (The TPT with loops and rings reduced to first-order TPT when those motifs were neglected.) Complete details are given in Ref.~\citenum{Howard:2020}, so we will summarize only the key results needed for this work. Some of the notation used here differs slightly, so we also quote the relevant equations from Ref.~\citenum{Howard:2020} in [square brackets] to facilitate comparison.

The fraction of linkers in bridge and loop motifs were computed from the chemical equilibrium equations [Eqs.~(57), (58), (68), and (69)],
\begin{align}
-1 + \frac{X_{\Gamma_{\rm c}\setminus A_1}^{({\rm c})}}{X_{\Gamma_{\rm c}}^{({\rm c})}} &= 2 \rho_{\rm l} X_{B_1}^{({\rm l})} \Delta_{\rm cl}
\label{eq:eq1c} \\
-1 + \frac{X_{B_1}^{({\rm l})}}{X_{B_{12}}^{({\rm l})}} &= n_{\rm c} \rho_{\rm c} X_{A_1}^{({\rm c})} \Delta_{\rm cl}
\label{eq:eq1p} \\
\frac{X_{\Gamma_{\rm c}\setminus A_{12}}^{({\rm c})}}{X_{\Gamma_{\rm c}}^{({\rm c})}} - \left(\frac{X_{\Gamma_{\rm c}\setminus A_1}^{({\rm c})}}{X_{\Gamma_{\rm c}}^{({\rm c})}}\right)^2 &= 2 \rho_{\rm l} X_{B_{12}}^{({\rm l})} \Delta_2 +16 \bigg(\frac{\nu_{\rm c}}{2}\bigg) \bigg(\rho_{\rm c} X_{A_{12}}^{({\rm c})}\bigg) \bigg(\rho_{\rm l} X_{B_{12}}^{({\rm l})}\bigg)^2 \Delta_4
\label{eq:eq4c} \\
\frac{1}{X_{B_{12}}^{({\rm l})}} - \left(\frac{X_{B_1}^{({\rm l})}}{X_{B_{12}}^{({\rm l})}}\right)^2 &= 2 \nu_{\rm c} \rho_{\rm c} X_{A_{12}}^{({\rm c})} \Delta_2 + 16 \bigg(\frac{\nu_{\rm c}^2}{2}\bigg) \bigg(\rho_{\rm c} X_{A_{12}}^{({\rm c})}\bigg)^2 \bigg(\rho_{\rm l} X_{B_{12}}^{({\rm l})}\bigg) \Delta_4.
\label{eq:eq4p}
\end{align}
$X_\alpha^{(i)}$ is the fraction of component $i$ that is not bonded at a particular set of sites $\alpha$, $\Gamma_{\rm c}$ is the set of six bonding sites on the colloids, $A_{12} = \{A_1,A_2\}$ is the set of two (representative) neighboring bonding sites on the colloid, and $B_{12} = \{B_1,B_2\}$ is the set of bonding sites on the linker. Due to the colloid bonding site geometry in our model, there are $\nu_{\rm c} = n_{\rm c}(n_{\rm c}-2)/2 = 12$ equivalent pairs of colloid bonding sites that can participate in a loop or ring. Inclusion of the loop and ring structures in the TPT introduces two bond volumes $\Delta_2$ and $\Delta_4$ in addition to $\Delta_{\rm cl}$ (Eq.~(12) of the main text, $\Delta_1$ [Eq.~(61)] in Ref.~\citenum{Howard:2020}). The loop (double bond) volume $\Delta_2$ is [Eq.~(62)]
\begin{align}
\Delta_2 \approx \frac{1}{8\pi^2} \int \dd\vv{r} \dd\vvg{\Omega} \dd\vv{R} \bigg[p_0(\vv{R}) g_0^{({\rm cl})}(\vv{r},\vv{R}) f_{A_1 B_1}^{({\rm cl})}(\vv{r},\vvg{\Omega}) f_{A_2 B_2}^{({\rm cl})}(\vv{r},\vvg{\Omega},\vv{R}) \bigg],
\label{eq:delta4}
\end{align}
which is similar to $\Delta_{\rm cl}$ but has a nonzero integrand only when the pairs $A_1$--$B_1$ and $A_2$--$B_2$ are both interacting. (The subscripts of the $f$-functions denote the interacting pairs of patches.) $\Delta_4$ is the ring volume that requires interactions between four pairs of bonding sites and is given by Eq.~(70) of Ref.~\citenum{Howard:2020}. To close the system of nonlinear equations, the values of $X_\alpha^{({\rm c})}$ are related by [Eqs.~(B11), (B12), and (B13)]
\begin{align}
1 = &\bigg[11 \Big(X_{\Gamma_{\rm c}\setminus A_1}^{({\rm c})}\Big)^6 - 24 X_{\Gamma_{\rm c}\setminus A_{12}}^{({\rm c})} \Big(X_{\Gamma_{\rm c}\setminus A_1}^{({\rm c})}\Big)^4 X_{\Gamma_{\rm c}}^{({\rm c})} + 6 \Big(X_{\Gamma_{\rm c}\setminus A_{12}}^{({\rm c})}\Big)^2 \Big(X_{\Gamma_{\rm c}\setminus A_1}^{({\rm c})}\Big)^2 \Big(X_{\Gamma_{\rm c}}^{({\rm c})}\Big)^2 \nonumber \\
&+ 8 \Big(X_{\Gamma_{\rm c}\setminus A_{12}}^{({\rm c})}\Big)^3 \Big(X_{\Gamma_{\rm c}}^{({\rm c})}\Big)^3\bigg]\Big/\Big(X_{\Gamma_{\rm c}}^{({\rm c})}\Big)^5 \\
X_{A_1}^{({\rm c})} = &\bigg[3 \Big(X_{\Gamma_{\rm c}\setminus A_1}^{({\rm c})}\Big)^5 - 12 X_{\Gamma_{\rm c}\setminus A_{12}}^{({\rm c})} \Big(X_{\Gamma_{\rm c}\setminus A_1}^{({\rm c})}\Big)^3 X_{\Gamma_{\rm c}}^{({\rm c})} \nonumber \\ 
&+ 10 \Big(X_{\Gamma_{\rm c}\setminus A_{12}}^{({\rm c})}\Big)^2 X_{\Gamma_{\rm c}\setminus A_1}^{({\rm c})} \Big(X_{\Gamma_{\rm c}}^{({\rm c})}\Big)^2\bigg]\Big/\Big(X_{\Gamma_{\rm c}}^{({\rm c})}\Big)^4 \\
X_{A_{12}}^{({\rm c})} = &\bigg[-2 \Big(X_{\Gamma_{\rm c}\setminus A_1}^{({\rm c})}\Big)^4 + X_{\Gamma_{\rm c}\setminus A_{12}}^{({\rm c})} \Big(X_{\Gamma_{\rm c}\setminus A_1}^{({\rm c})}\Big)^2 X_{\Gamma_{\rm c}}^{({\rm c})} + 2 \Big(X_{\Gamma_{\rm c}\setminus A_{12}}^{({\rm c})}\Big)^2 \Big(X_{\Gamma_{\rm c}}^{({\rm c})}\Big)^2\bigg]\Big/\Big(X_{\Gamma_{\rm c}}^{({\rm c})}\Big)^3.
\end{align}
We numerically solved \cite{numpy,scipy,numba} these equations for the five independent unknowns $X_{\Gamma_{\rm c}\setminus A_{12}}^{({\rm c})}$, $X_{\Gamma_{\rm c}\setminus A_1}^{({\rm c})}$, $X_{\Gamma_{\rm c}}^{({\rm c})}$, $X_{B_1}^{({\rm l})}$, and $X_{B_{12}}^{({\rm l})}$ using the procedure described in Appendix B of Ref.~\citenum{Howard:2020}. From these variables, it follows that the fraction of linkers bonded in loops $\chi_{\rm loop}$ was [Eq.~(71)]
\begin{equation}
\chi_{\rm loop} = 2\nu_{\rm c} \rho_{\rm c} X_{A_{12}}^{({\rm c})} X_{B_{12}}^{({\rm l})} \Delta_2,
\end{equation}
the fraction of linkers bonded at both ends $\chi_{B_{12}}$ was [Eq.~(13)]
\begin{equation}
\chi_{B_{12}} = 1 - 2 X_{B_1}^{({\rm l})} + X_{B_{12}}^{({\rm l})},
\end{equation}
and the bridge fraction $\chi_{\rm bridge}$ was
\begin{equation}
\chi_{\rm bridge} = \chi_{B_{12}} - \chi_{\rm loop}.
\end{equation}

We evaluated $\Delta_{\rm cl}$, $\Delta_2$, and $\Delta_4$ using the approach and approximations detailed in Appendix B of Ref.~\citenum{Howard:2020} [Eqs.~(B3), (B7), and (B10)]. We first collected reference fluid simulation data for $p_0$ and $g_0^{({\rm cl})}$ using the methodology described in Ref.~\citenum{Howard:2020}; the histogram bin size $\Delta R$ for the end-to-end distance was adjusted for each bending stiffness as listed in Table~\ref{tab:fits}. We fit the simulated distribution of end-to-end vectors $p_0(\vv{R})$ to a piecewise stretched-exponential form,
\begin{equation}
p_0(R) = \begin{cases}
p_{\rm max} \exp[-c_1 (R_{\rm max}-R)^{c_2}],& R \le R_{\rm max} \\
p_{\rm max} \exp[-c_3(R-R_{\rm max})^{c_4}],& R > R_{\rm max} \\
\end{cases}
\end{equation}
with fitting parameters $p_{\rm max}$, $R_{\rm max}$, $c_1$, $c_2$, $c_3$, and $c_4$; the resulting curves are shown in Figure 3.

\begin{figure}
    \centering
    \includegraphics{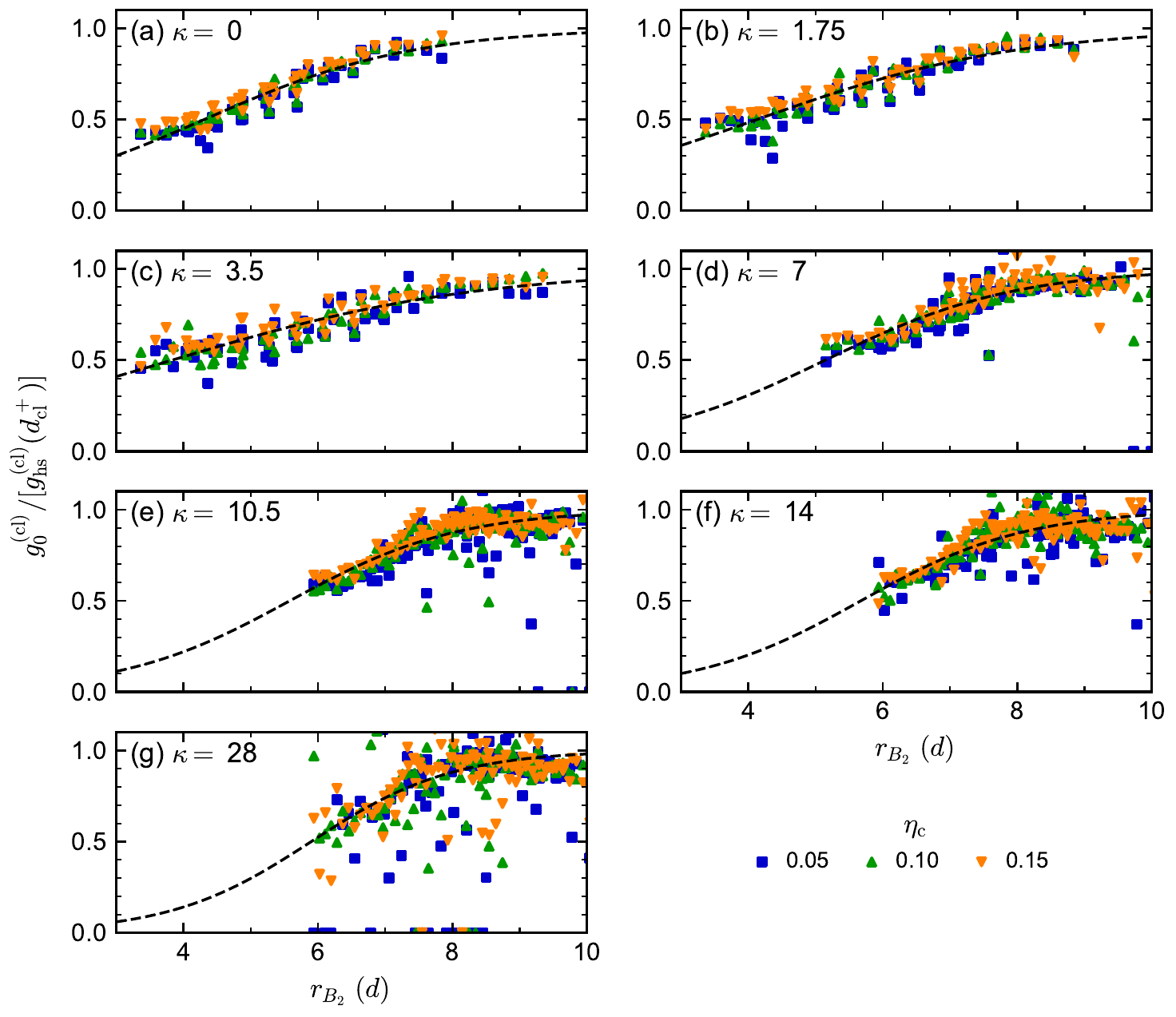}
    \caption{Simulated colloid--linker pair distribution function $g_0^{({\rm cl})}$ in reference mixtures having varied $\kappa$, varied $\eta_{\rm c}$, and fixed $\Gamma = 1.5$. $g_0^{({\rm cl})}$ is normalized by the contact value of the pair distribution function $g_{\rm hs}^{({\rm cl})}(d_{\rm cl}^+)$ in the hard-sphere mixture that would be obtained by removing all bonds from the linkers \cite{Boublik:1970}. One linker end is fixed at distance $r_{B_1} = d_{\rm cl}^*$ from the center of the colloid, and $g_0^{({\rm cl})}/[g_{\rm hs}^{({\rm cl})}(d_{\rm cl}^+)]$ collapses as a function (dashed line) of the distance from the colloid center to the other linker end $r_{B_2}$. Here, we show only data only for which we have reliable sampling, with ranges listed in Table \ref{tab:fits}.}
    \label{fig:gsb}
\end{figure}

The values of $g_0^{({\rm cl})}$ relevant for evaluating $\Delta_{\rm cl}$ [Eq.~(B3)] and $\Delta_4$ [Eq.~(B10)] involve linker conformations where one end is at a distance $r_{B_1} = d_{\rm cl}^*$ from the center of the colloid (i.e., near the colloid bonding site). The volume-fraction dependence of $g_0^{({\rm cl})}$ was accounted for using the contact value of the pair distribution function $g_{\rm hs}^{({\rm cl})}(d_{\rm cl}^+)$ in the hard-sphere mixture that would be obtained by removing all bonds from the linkers \cite{Boublik:1970}, making the data a function of only the distance $r_{B_2}$ from the center of the colloid to the other linker end and that was well fit using an S-curve (Figure~\ref{fig:gsb}),
\begin{equation}
    g_0^{({\rm cl})}(d_{\rm cl}^*,r_{B_2}) \approx \frac{g_{\rm hs}^{({\rm cl})}(d_{\rm cl}^+)}{1+ C_1 \exp(-C_2 r_{B_2})},
    \label{eq:g0sbfit}
\end{equation}
with fitting parameters $C_1$ and $C_2$ (Table~\ref{tab:fits}). We used these fits, with extrapolation, to evaluate $\Delta_{\rm cl}$ and $\Delta_4$.

\begin{figure}
    \centering
    \includegraphics{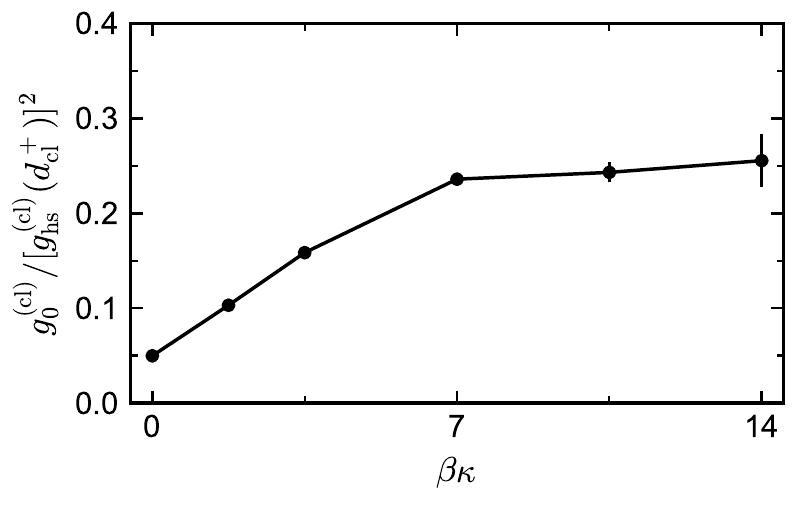}
    \caption{Simulated colloid--linker pair distribution function $g_0^{({\rm cl})}$ when both ends of the linker are fixed at colloid bonding sites so that $r_{B_1} = d_{\rm cl}^*$, $R = \sqrt{2} d_{\rm cl}^*$, and $\phi = 3\pi/4$. The value shown is the average for colloid volume fractions $0.05 \le \eta_{\rm c} \le 0.15$ (in steps of 0.01), and the error bars indicate one standard error of the mean.}
    \label{fig:gdb}
\end{figure}

The values of $g_0^{({\rm cl})}$ relevant for evaluating $\Delta_2$ [Eq.~(B7)] involve linker conformations where both ends are near neighboring colloid bonding sites. The volume-fraction dependence of this value was accounted for using a factor of $[g_{\rm hs}^{({\rm cl})}(d_{\rm cl}^+)]^2$. The proportionality constant $C_3 = g_0^{({\rm cl})}/[g_{\rm hs}^{({\rm cl})}(d_{\rm cl}^+)]^2$ is shown in Figure~\ref{fig:gdb} and listed in Table~\ref{tab:fits} for $\beta\kappa \le 14$. We did not have reliable statistics for $\beta\kappa = 28$ because such linker conformations were highly improbable; we accordingly used $C_3$ for $\beta\kappa = 14$ in our calculations for $\beta\kappa = 28$.

We note that we also attempted to compute the spinodal boundary using the free-energy density $a$ given by the TPT \cite{Howard:2020}. Due to the computational expense of evaluating the equilibrium bonding states of the colloids and linkers for the association free-energy density $a_{\rm b}$, we numerically tabulated $a_{\rm b}$ as a function of composition, interpolated it using bivariate spline functions, and differentiated to obtain the stability matrix $\vv{H}$. We encountered significant numerical issues with this approach, which we also encountered using $a_{\rm b}$ given by first-order TPT, and these difficulties prevented us from reliably identifying the spinodal boundary. We suspect these difficulties are caused by insufficient precision and accuracy determining $X_\alpha^{(i)}$ in the limit of large $\varepsilon$. Given the good accuracy of the TPT with loops and rings for computing the various bonding states of the linkers (Figure 4), it would be beneficial to identify more accurate numerical schemes for computing the spinodal boundary, as this is important for guiding experiments that realize linked colloidal gels.

\clearpage
\begin{table}
    \centering
    \caption{Fitting ranges and parameters for $g_0^{({\rm cl})}$ at different bending stiffnesses $\kappa$. The linker conformation was defined by its end-to-end vector $\vv{R}$ from $B_1$ to $B_2$ in spherical coordinates, using $B_1$ as the origin and the vector from the colloid center to $B_1$ as the axis. The relevant coordinates for fitting are the linker end-to-end distance $R$ and polar angle $\phi$. Fitting was restricted to the probable ranges of $R$ based on Figure 3 and to $\phi \le \pi/2$ based on steric considerations. $\Delta R$ is the histogram bin width for the end-to-end distance used to compute $p_0$ and $g_0^{({\rm cl})}$ in the simulations, $C_1$ and $C_2$ are the fitting parameters in Eq.~\eqref{eq:g0sbfit}, and $C_3$ is the proportionality constant in Figure~\ref{fig:gdb}.}
    \begin{tabular}{c c c c c c c}
         $\beta \kappa$ & $\Delta R/d$ & min $R/d$ & max $R/d$ & $C_1$ & $C_2$ & $C_3$ \\ \hline
         0 & 0.5 & 1 & 5 & 16.0 & 0.642 & 0.0496 \\
         1.75 & 0.5 & 1 & 6 & 8.52 & 0.518 & 0.103 \\
         3.5 & 0.5 & 1 & 6.5 & 5.34 & 0.437 & 0.158 \\
         7 & 0.2 & 4 & 7 & 38.3 & 0.709 & 0.236 \\
         10.5 & 0.1 & 5 & 7 & 86.1 & 0.798 & 0.243 \\
         14 & 0.1 & 5 & 7 & 103. & 0.827 & 0.256 \\
         28 & 0.1 & 5 & 7 & 275. & 0.951 & $-$
    \end{tabular}
    \label{tab:fits}
\end{table}

\begin{figure}
    \centering
    \includegraphics{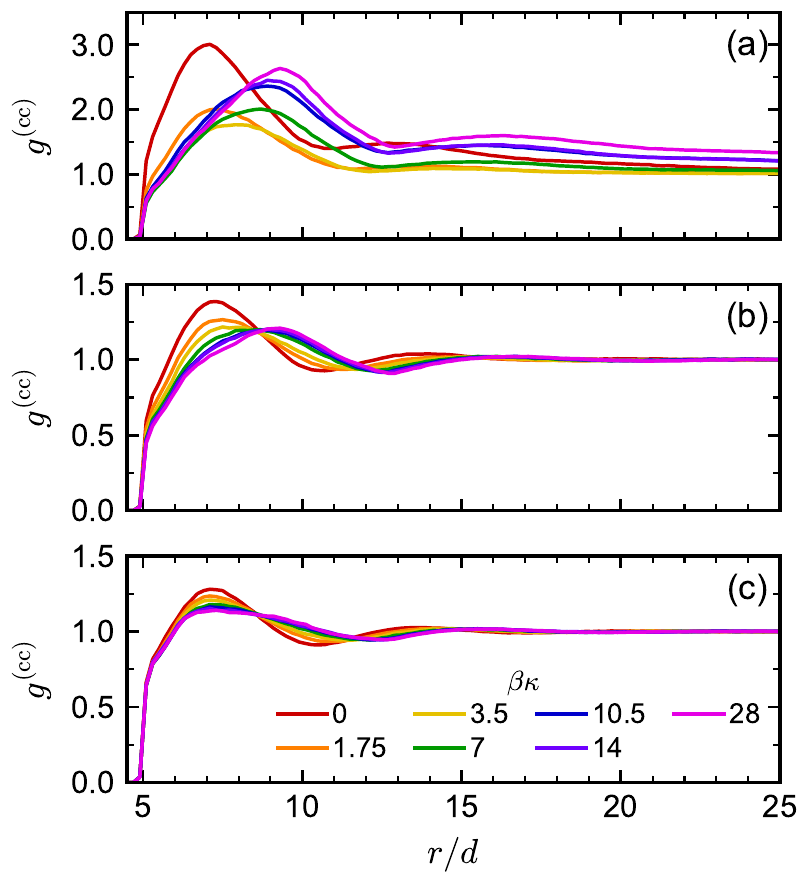}
    \caption{Colloid radial distribution function $g_{\rm cc}$ for different bending stiffnesses $\kappa$ at $\beta\varepsilon = 20$, $\Gamma = 1.5$, and (a) $\eta_{\rm c} = 0.03$, (b) 0.10, and (c) 0.15.}
    \label{fig:rdf}
\end{figure}

\clearpage
\bibliography{ms}